\documentclass[%
reprint,
amsmath,amssymb,
aps,
prb,
]{revtex4-1}

\usepackage{graphicx}
\usepackage{dcolumn}
\usepackage{bm}

\newcommand*{\citen}{}
\DeclareRobustCommand*{\citen}[1]{%
	\begingroup
	\romannumeral-`\x 
	\setcitestyle{numbers}%
	\cite{#1}%
	\endgroup
}

\begin{document}
	
\preprint{APS/123-QED}
	
	
\title{Hidden Order Signatures in the Antiferromagnetic Phase of U(Ru$_{1-x}$Fe$_x$)$_2$Si$_2$}
	
\author{T.J.~Williams}
\email{williamstj@ornl.gov}
\affiliation{Quantum Condensed Matter Division,
	Neutron Sciences Directorate,
	Oak Ridge National Lab,
	Oak Ridge, TN, 37831, USA}
	
\author{M.N.~Wilson}
\affiliation{Department of Physics and Astronomy,
	McMaster University,
	Hamilton, ON, L8S 4M1, Canada}
	
\author{A.A.~Aczel}
\author{M.B.~Stone}
\affiliation{Quantum Condensed Matter Division,
	Neutron Sciences Directorate,
	Oak Ridge National Lab,
	Oak Ridge, TN, 37831, USA}

\author{G.M.~Luke}
\affiliation{Department of Physics and Astronomy,
	McMaster University,
	Hamilton, ON, L8S 4M1, Canada}
\affiliation{Canadian Institute for Advanced Research,
	180 Dundas St. W., 
	Toronto, ON, M5G 1Z7, Canada}

\date{\today}

\begin{abstract}

We present a comprehensive set of elastic and inelastic neutron scattering 
measurements on a range of Fe-doped samples of U(Ru$_{1-x}$Fe$_x$)$_2$Si$_2$ 
with 0.01~$\le~x~\le$~0.15.  All of the samples measured exhibit long-range 
antiferromagnetic order, with the size of the magnetic moment quickly 
increasing to 0.51~$\mu_B$ at 2.5\% doping and continuing to increase 
monotonically with doping, reaching 0.69~$\mu_B$ at 15\% doping.  
Time-of-flight and inelastic triple-axis measurements show the existence of 
excitations at (1~0~0) and (1.4~0~0) in all samples, which are also observed 
in the parent compound.  While the excitations in the 1\% doping are 
quantitatively identical to the parent material, the gap and width of the 
excitations change rapidly at 2.5\% Fe doping and above.  The 1\% doped sample 
shows evidence for a separation in temperature between the hidden order and 
antiferromagnetic transitions, suggesting that the antiferromagnetic state 
emerges at very low Fe dopings.  The combined neutron scattering data suggests 
not only discontinuous changes in the magnetic moment and excitations between 
the hidden order and antiferromagnetic phases, but that these changes continue 
to evolve up to at least $x$~=~0.15.
	
\begin{description}
\item[PACS numbers]{71.27.+a, 75.25.-j, 75.40.Gb, 78.70.Nx}
\end{description}
\end{abstract}
	
\maketitle
	
\section{\label{sec:level1}Introduction}
	
The heavy fermion material URu$_2$Si$_2$ has been a subject of long-standing 
interest since the discovery of a phase transition at T$_0$~=~17.5~K, thirty 
years ago~\cite{Palstra_85}.  Initially thought to be an antiferromagnetic 
transition, the small antiferromagnetic moment of 0.03~$\mu_B$ that arises in 
this material is far too small to account for the large specific heat jump at 
T$_0$~\cite{Broholm_87,Broholm_91}.  Three decades of research have produced a 
number of conclusions regarding the nature of this 
phase~\cite{Mydosh_11,Mydosh_14}, but have failed to determine the order 
parameter, leading to this phase being dubbed the `hidden order' phase.  
To study the behavior of the hidden order phase, a large number 
of perturbations have been applied to the system in the form of applied field, 
hydrostatic pressure and chemical substitution.  In all cases, the hidden 
order phase is destroyed with relatively small perturbations: applied fields 
of $>$35~T~\cite{Jo_08}, hydrostatic pressure $>$0.8~GPa~\cite{Butch_10} and 
chemical substitution of typically greater than 5~\% on any of the atomic 
sites~\cite{Amitsuka_94,Endstra_93,Park_93}.  In nearly every case, the hidden 
order state is suppressed continuously, and a ferro- or antiferromagnetic 
state emerges.  

Neutron scattering has played an important role in determining the properties 
of the hidden order phase.  For example, while careful study has shown that 
the small antiferromagnetic moment is present even in ultra-clean 
samples~\cite{Bourdarot_14}, it is likely caused by inhomogeneous 
strain~\cite{Amitsuka_07}.  Within the paramagnetic phase above T$_0$, 
inelastic neutron scattering measurements observed gapless, weakly dispersing 
features at the $\Sigma$ point on the Brillouin Zone (BZ) edge with 
$\vec{Q}_{inc}$~=~(1$\pm \delta$~0~0) ($\delta$~=~0.407), while below T$_0$, 
these excitations became gapped 
($\Delta_{inc}$~=~4.5-4.8~meV~\cite{Bourdarot_14,Butch_15}) and more 
intense~\cite{Broholm_91,Wiebe_07}.  It was determined that the gapping of 
these excitations results in an entropy change of sufficient size to account 
for the specific heat jump at T$_0$~\cite{Wiebe_07}.  Below T$_0$ additional, 
commensurate excitations appear at the $Z$ point of the BZ, 
$\vec{Q}_{com}~=~(1~0~0)$, with a gap of 
$\Delta_{com}$~=~1.7-1.8~meV~\cite{Bourdarot_14,Butch_15}.  This wavevector is 
the ordering wavevector for the antiferromagnetic moment in both the hidden 
order and more conventional magnetically-ordered phases.  Since the transition 
at T$_0$ is related to the gapping of the incommensurate excitations and the 
emergence of the commensurate ones, these have both been cited as possible 
`signatures' of the hidden order state in neutron scattering 
experiments~\cite{Mydosh_14,Bourdarot_14}.

The first instance in which perturbations were found to enhance the hidden 
order state was through the use of applied pressure.  Application of pressure 
increased T$_0$ slightly, reaching 18.5~K at a pressure of 
0.5~GPa~\cite{Butch_10}.  However, at higher pressures, this system still 
transitions to an antiferromagnetic state; at T~=~0 this occurs at 
approximately 0.8~GPa.  Pressures between 0.8 and 1.4~GPa have both a hidden 
order and a Ne\'{e}l transition, while above 1.4~GPa the transition is 
directly from paramagnetic to antiferromagnetic at 
T$_N$~=~19.5~K~\cite{Butch_10}.  Due to this interplay of hidden order and 
antiferromagnetism, studying the behavior under applied pressure has 
become of particular interest in trying to determine the nature of the unknown 
order parameter.  Likewise, the chemical substituents that enhance T$_0$ have 
also become an interesting avenue of research for determining the order 
parameter of the hidden order state.  Of the dozens of chemical dopings that 
have been applied to URu$_2$Si$_2$ only two dopings, both on the Ru site, have 
been shown to increase the value of T$_0$: Fe~\cite{Kanchanavatee_11} and 
Os~\cite{Kanchanavatee_14}.  In both of these cases, the transition 
temperature continues to increase as a function of doping, over a large range, 
before dropping abruptly.  Interestingly, of all of the pure compounds of the 
family U$T_2$Si$_2$, $T$~=~Fe and Os are the only two that are 
non-magnetic~\cite{Endstra_93,Sandratskii_94}.  Furthering the analogy between 
hydrostatic pressure and Fe/Os-doping, the doped systems are also observed to 
become more conventionally antiferromagnetic with increasing chemical 
pressure, however no signature of multiple transitions have been observed with 
transport measurements~\cite{Kanchanavatee_11,Kanchanavatee_14}.  It was 
speculated that these systems experience only a gradual crossover between the 
hidden order and antiferromagnetic states, although this remains an open 
question.

In this work, we use elastic and inelastic neutron scattering to measure the 
magnetic structure and excitations of various doping concentrations within the 
U(Ru$_{1-x}$Fe$_x$)$_2$Si$_2$ series, in an attempt to determine the nature of 
the hidden order-to-antiferromagnetic crossover, as well as whether the doped 
compounds contain inelastic signatures of the hidden order state and/or 
signatures of a conventional antiferromagnetic state (spin waves).  Recently, 
neutron diffraction measurements have been carried out on a number of dopings 
in this series~\cite{Das_15}, which found that the magnetic moment grows 
continuously from $x$~=~0 to $x$~=~0.05 and that at dopings above 5\% the 
magnetic moment remains relatively constant at 0.8~$\mu_B$.  This leads the 
authors to suggest that 5\% doping marks the hidden order-to-antiferromagnetic 
phase transition, analogous to the transition at 0.8~GPa in the parent 
compound under pressure~\cite{Das_15}.  This suggests that in order to study 
the nature of the excitations through the transition, it is important to 
measure dopings both above and below $x$~=~0.05.

\section{\label{sec:level2}Experiment}

Single crystals of U(Ru$_{1-x}$Fe$_x$)$_2$Si$_2$ with $x$~=~0.01, 0.025, 0.05, 
0.10 and 0.15 were grown at McMaster University.  Stoichiometric amounts of 
unpurified depleted Uranium, Ru (99.95\%), Fe (99.99\%) and Si (99.9999\%) 
were arc-melted on a water-cooled copper hearth in a mono-arc furnace under an 
inert Ar atmosphere.  The largest impurity in the Uranium precursor is 
elemental Fe at a level of $\approx$50 ppm, which is small ($<$0.01\%) when 
compared to the nominal doping concentrations.  The resulting polycrystalline 
boule was then used to grow the single crystals using the Czochralski method.  
This was performed in a tri-arc furnace using a water-cooled copper hearth 
under a continuously-gettered Ar atmosphere at 900~$^{\circ}$C.  After the 
growths, the single-crystalline nature and sample alignments were confirmed 
with Laue x-ray diffraction.
	
These samples were studied using elastic and inelastic neutron scattering at 
the High-Flux Isotope Reactor (HFIR) and the Spallation Neutron Source (SNS) 
of Oak Ridge National Laboratory (ORNL).  The diffraction measurements were 
performed on all of the samples using the HB-1A spectrometer at HFIR, while 
inelastic measurements were done on the HB-1 (for $x$~=~0.01 and 0.05) and 
HB-3 (for $x$~=0.025, 0.10 and 0.15) triple-axis instruments at HFIR, as well 
as the SEQUOIA time-of-flight spectrometer at the SNS (for $x$=0.05 and 
0.15).  For comparison, data on the parent compound has been included where 
appropriate; this data was measured on the Multi-Axis Crystal Spectrometer 
(MACS) at the NIST Center for Neutron Research and was published 
previously~\cite{Williams_16}.  The neutron measurements described in this 
work were performed using 1 single crystal of each doping: the $x$~=~0.01 
sample had a mass of 5.65(2)~g and a mosaic of 4.5$^{\circ}$; the $x$~=~0.025 
sample had a mass of 1.99(1)~g and a mosaic of 1.3$^{\circ}$; the $x$~=~0.05 
sample had a mass of 2.98(1)~g and a mosaic of 10$^{\circ}$; the $x$~=~0.10 
sample had a mass of 1.85(1)~g and a mosaic of 3.0$^{\circ}$; and the 
$x$~=~0.15 sample had a mass of 1.74(1)~g and a mosaic of 4.0$^{\circ}$.  All 
of these samples were aligned in the [H~0~L] scattering plane for each of the 
neutron scattering experiments.


The HB-1A measurements were performed in a closed-cycle refrigerator with a 
base temperature of 4.0~K using a fixed incident energy of 14.7~meV.  PG (002) 
monochromator and analyzer crystals were used with PG filters, and the 
collimation was 40'-40'-40'-80'.  The HB-1 and HB-3 measurements were 
performed in closed-cycle refrigerators with a base temperature of 4.0~K using 
a fixed final energy of 14.7~meV.  PG (002) monochromator and analyzer 
crystals were used with PG filters, and the collimation was 48'-40'-40'-120'.  
The SEQUOIA measurements were also performed in a closed-cycle refrigerator 
with a base temperature of 5~K, using a fixed incident energy of 30~meV.  The 
crystals were rotated in the [H~0~L] plane in 1$^{\circ}$ steps over a 
190$^{\circ}$ range.

\section{\label{sec:level3}Magnetic Structure Determination}

The neutron diffraction involved measurements of all of the Bragg peaks for 
which $|\vec{Q}|$~$<$~4.7~\AA$^{-1}$, at 4~K and 30~K, as well as the 
temperature dependence of the (1~0~0) and (0~0~1) magnetic Bragg peaks.  While 
the (0~0~1) peaks was found to have a weak magnetic signal, the $c$-axis 
magnetic contribution was found to be consistent with what would be expected 
due to multiple scattering for E$_i$~=~14.7~meV, suggesting that the magnetic 
moments point along the $\hat{c}$-direction.  Multiple scattering was also 
encountered in the parent material, where the same magnetic structure was 
refined for the small, intrinsic moments~\cite{Ross_14}.

\begin{figure}[tbh]
\begin{center}
\includegraphics[angle=0,width=\columnwidth]{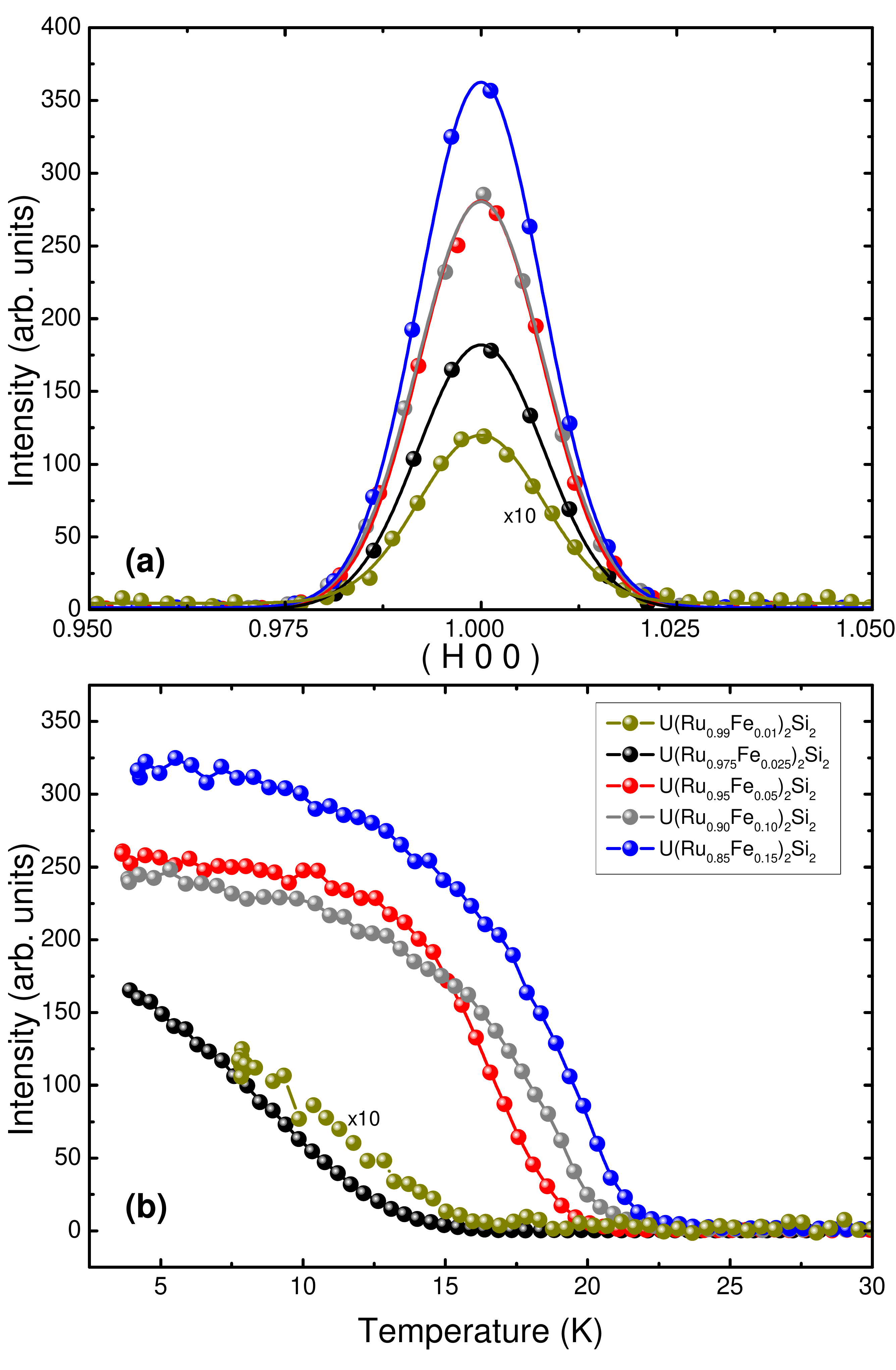}
\caption{\label{hb1a} (color online) (a) Radial scans through the (1~0~0) 
magnetic Bragg peaks at T~=~4~K in the various samples of 
U(Ru$_{1-x}$Fe$_x$)$_2$Si$_2$.  All of the peaks appear resolution-limited, 
indicating long-range magnetic order.  This is a disallowed nuclear peak, and 
so there is no scattering from the sample above T$_0$.  (b) The 
temperature-dependence of the (1~0~0) magnetic Bragg peak intensity in the 
various samples.  This shows the second-order transition from the paramagnetic 
state to the antiferromagnetic state at T$_N$.  The lack of saturation of the 
moment in the 1\% (yellow) and 2.5\% (black) samples may be dues to phase 
separation (see text).  In both plots, the error bars lie within the symbols.}
\end{center}
\end{figure}

Fig.~\ref{hb1a} shows the (1~0~0) magnetic Bragg peak at 4~K in the various 
Fe-doped samples (panel (a)) and their temperature dependence (panel (b)).  
This is a disallowed nuclear peak so there is no scattering from the sample 
above T$_0$, as seen in the temperature-dependence.  We observe the onset of 
magnetic scattering, and the transition appears to be second order in nature.  
The temperature dependence of the lowest two dopings, 1\% and 2.5\% do not 
show the same temperature dependence.  Previous work using $\mu$SR has shown 
that at these dopings, there is considerable phase separation between magnetic 
and non-magnetic regions, likely as a result of the random dopant distribution 
in these samples~\cite{Wilson_16}.  This is a likely origin of the observed 
temperature dependence of the magnetic Bragg peak.  However, the peaks are 
resolution-limited at all dopings, suggesting that the magnetic order is 
sufficiently long-ranged.  Using the 7 structural and 9 magnetic peaks 
collected on each sample, the magnetic structure and moment can be determined. 
In agreement with the parent material at ambient pressure and in the 
pressure-induced antiferromagnetic state, this magnetic structure has magnetic 
moments aligned along the $c$-axis, with the body-centered moment antiparallel 
to the moments in the neighboring $ab$-planes~\cite{Ross_14}.

\begin{table}[htb]
\begin{center}
\begin{tabular}{|c|l|l|l|l|}
\hline
Doping (\%) & ~T$_N$ (K)~ & Moment ($\mu_B$) & ~T$_0$ (K)~\cite{Wilson_16}~ & ~T$_N$ (K)~\cite{Wilson_16}~
\\
\hline
~1.0~\% & ~15.0(5) & ~~~0.048(5) & ~~~17.5 & ~~~16.0  \\
~2.5~\% & ~15.0(5) & ~~~0.51(1)  & ~~~     & ~~~      \\
~5.0~\% & ~20.0(5) & ~~~0.59(1)  & ~~~21.0 & ~~~21.0  \\
10.0~\% & ~21.0(5) & ~~~0.59(2)  & ~~~21.5 & ~~~21.0  \\
15.0~\% & ~22.5(5) & ~~~0.66(2)  & ~~~25.5 & ~~~25.0  \\
\hline
\end{tabular}
\end{center}
\caption[]{The transition temperatures and extracted moment sizes in the 
various dopings of U(Ru$_{2-x}$Fe$_x$)$_2$Si$_2$ measured in this work.  The 
value of T$_N$ is the transition temperature seen in the measurement of the 
(1~0~0) Bragg peak (Fig.~\ref{hb1a}(b)).  Also listed are the values of T$_0$ 
and T$_N$ as determined from the same crystals that were used in the current 
studies.  These values were obtained from susceptibility and $\mu$SR 
measurements as reported in Ref.~\citen{Wilson_16}.}
\label{moment_tbl}
\end{table}

The magnetic moment as a function of doping at T~=~4~K was extracted from the 
integrated intensity of the (1~0~0) magnetic peak normalized by the integrated 
intensity of the (1~0~1) structural peak, with the proper Lorentz factors 
taken into account for both Bragg peaks. The (1~0~1) structural peak was 
chosen for the normalization to minimize the difference in instrumental 
Q-resolution at the two peak positions, since resolution effects were not 
incorporated in these calculations. This approach is in contrast to the method 
employed by Das {\em et al}~\cite{Das_15}, who chose the higher order Bragg 
peak (6~0~0) for the normalization to avoid extinction effects.  Neither 
normalization method accounts for the effect of multiple-scattering, which has 
been noted as significant in URu$_2$Si$_2$, but that is difficult to calculate 
directly~\cite{Bourdarot_14,Ross_14}. This may produce differences in the size 
of the magnetic moments determined.

The moments that were extracted from the neutron diffraction measurements are 
shown in Table~\ref{moment_tbl}, along with the values of T$_N$ and T$_0$ from 
$\mu$SR in a previous work~\cite{Wilson_16}.  The values of T$_N$ from the 
measurement of the (1~0~0) magnetic Bragg peak are lower than those found by 
$\mu$SR, likely due to the local probe nature of the $\mu$SR measurements.  
The size of the moments agree well with the values determined from the 
internal field measurements based on the muon precession frequency, suggesting 
they are sensitive to the same magnetic ordering.  The size of the moment in 
the Fe-doped samples is compareable to what is seen in the pressure-induced 
antiferromagnetic state of the parent compound~\cite{Amitsuka_99}, except for 
the lowest doping (1\%).  In the lowest-doped sample, the size of the internal 
field determined by $\mu$SR would suggest a moment size of $\sim$0.45~$\mu_B$, 
however this was associated with a reduced volume fraction of $\sim$0.6 at 
T=~5~K~\cite{Wilson_16}.  The decreased moment seen by the neutron 
measurements is likely due to the phase separation between antiferromagnetism 
and the hidden order phase observed by the $\mu$SR measurements.  This would 
indicate that the transition from hidden order to antiferromagnetism occurs at 
a doping between 1\% and 2.5\%, lower than that suggested by Das {\em et 
al.}~\cite{Das_15}.  While we speculate that the difference in the moments may 
result from a different normalization method, the difference in the doping 
dependence may also be a result of differences in nominal and actual doping 
concentrations.  

	
\section{\label{sec:level4}Inelastic Measurements}
	
\begin{figure*}[tbh]
\begin{center}
\includegraphics[angle=0,width=\textwidth]{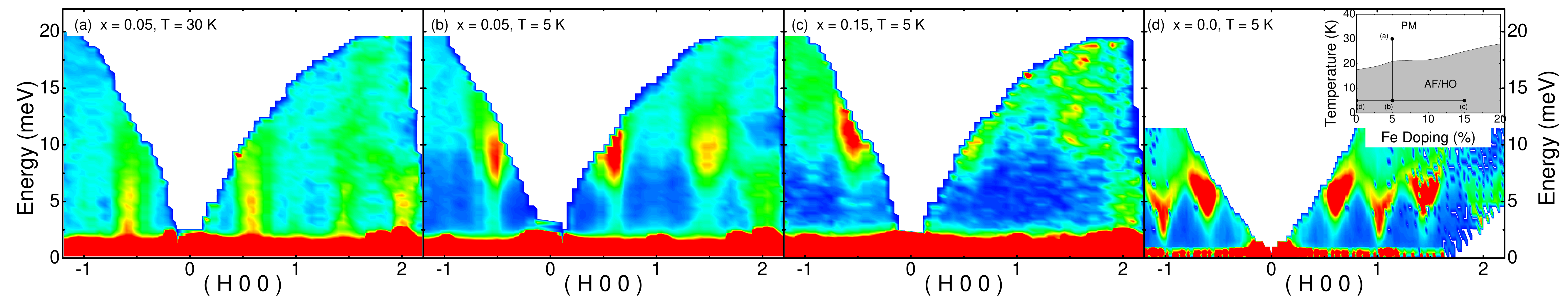}
\caption{\label{seq} (color online) Time-of-flight neutron measurements of 
various U(Ru$_{1-x}$Fe$_x$)$_2$Si$_2$ samples. (a) $x$~=~0.05, measured at 
30~K in the paramagnetic phase.  As is seen in the paramagnetic state of the 
parent ($x$~=~0) compound, there are gapless excitations at the incommensurate 
wavevector $\vec{Q}_{inc}$~=~(1.4~0~0). (b) Below T$_0$, these excitations 
become gapped and their spectral weight increases.  (c) At higher Fe dopings 
($x~=~0.15$ is shown here), the gap can be seen to increase and broaden in 
$\hbar \omega$ and $\vec{Q}$.  (d) Data from the parent compound (taken 
from Ref.~\citen{Williams_16}) below T$_0$ shows similar excitations at 
$\vec{Q}_{inc}$, however the excitations in the parent material are more 
well-defined.  Additionally, the commensurate excitations at 
$\vec{Q}_{com}$~=~(1~0~0), which are clearly present in the parent material, 
are not as obvious in the Fe-doped samples.  Cuts through $\vec{Q}_{com}$ show 
these excitations to be substantially weakened, and appear at higher energy 
than in the parent.  (inset) The phase diagram of 
U(Ru$_{1-x}$Fe$_x$)$_2$Si$_2$ showing the locations of the measurements for 
panels (a) to (d).}
\end{center}
\end{figure*}

Fig.~\ref{seq} shows the inelastic time-of-flight measurements of the 5\% 
sample at 30~K (panel (a)) and at 5~K (panel (b)), as well as the 15\% sample 
at 5~K (panel (c)).  Fig.~\ref{seq}(a) shows measurements in the paramagnetic 
state.  The inelastic spectrum seen here in the 5\%-doped sample is identical 
to what is seen in the parent material above T$_0$: gapless excitations 
emanating from $\vec{Q}_{inc}$~=~(0.6~0~0), and no excitation at 
$\vec{Q}_{com}$~=~(1~0~0).  Panel (d) illustrates what happens in the hidden 
order state of the parent material (this data is adapted from 
Ref.~\citen{Williams_16}).  The excitation at $\vec{Q}_{inc}$ becomes gapped, 
resulting in the entropy change seen by specific heat.  Additionally, gapped 
excitations also appear at $\vec{Q}_{com}$, albeit with a smaller gap and less 
intensity.  Fig.~\ref{seq}(b) shows the excitation spectrum below the 
transition in the 5.0\% Fe-doped sample.  Relative to the parent material, we 
see that the incommensurate excitation is qualitatively unchanged.  The gap 
appears to be larger, but with little change in the spin wave velocity, 
similar to what is observed under hydrostatic pressure~\cite{Williams_16}.  
The commensurate excitation, however, shows a large change when compared to 
the pure material in the hidden order state.  It is significantly weaker 
relative to the incommensurate excitation.  Furthermore, the scattering that 
is present at the commensurate point in the 5\% doping is only present at much 
higher energies.

Moving to higher Fe-doping (15.0\% in Fig.~\ref{seq}(c)), the weakening of 
these excitations seems to continue at both the commensurate and 
incommensurate points.  Additionally, we observe that the gap at 
$\vec{Q}_{inc}$ is larger than at $x$~=~0.05 or in the parent.  This type of 
trend has been observed under pressure, where an increase in the transition 
temperature seems to correlate with an increase in the incommensurate gap, 
though the magnitude of the gap change in this system is much larger than what 
has been observed under pressure for the same change in the transition 
temperature~\cite{Williams_16,Bourdarot_10}.  The excitations also appear 
broadened, both in $|Q|$ and $\hbar \omega$.  This would suggest that Fe 
doping distorts the Fermi surface, weakening the nesting that gives rise to 
the excitations~\cite{Butch_15}.  Furthermore, no additional excitations 
appear with Fe doping, including any conventional spin waves centered on the 
(1~0~0) magnetic Bragg peak.  To more carefully investigate the changes in the 
excitations, inelastic triple axis neutron scattering measurements were 
performed at both $\vec{Q}_{com}$ and $\vec{Q}_{inc}$, above and below T$_0$.

\begin{figure*}[tbh]
\begin{center}
\includegraphics[angle=0,width=\textwidth]{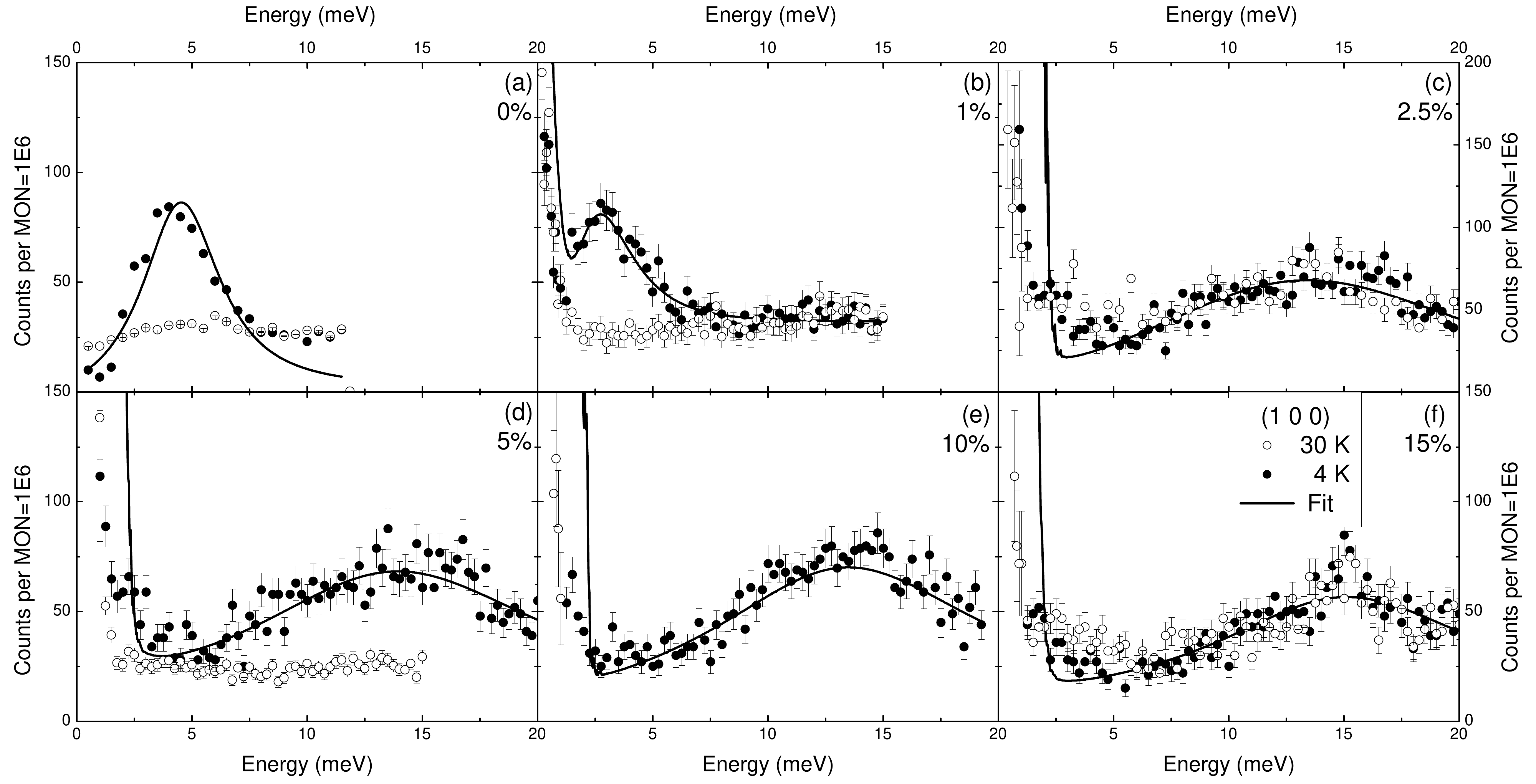}
\caption{\label{q_com} Commensurate excitation as a function of 
doping at T~=~4~K (filled circles) and T~=~30~K (open circles).  The solid 
line is a fit to the low temperature data as described in the text.  The data 
for the parent compound is adapted from Ref.~\citen{Williams_16}.}
\end{center}
\end{figure*} 

\begin{figure*}[tbh]
\begin{center}
\includegraphics[angle=0,width=\textwidth]{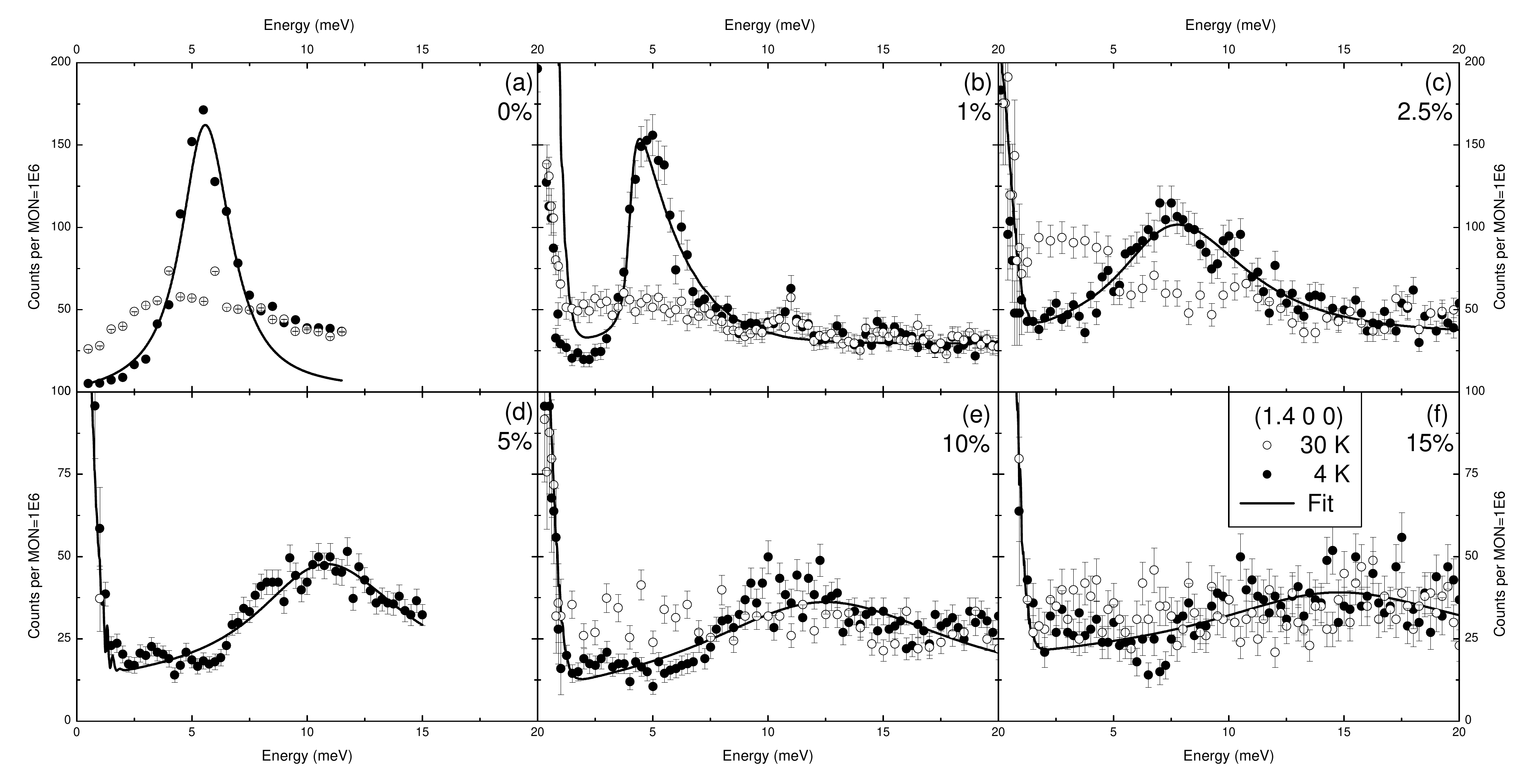}
\caption{\label{q_inc} Incommensurate excitation as a function 
of doping at T~=~4~K (filled circles) and T~=~30~K (open circles).  The solid 
line is a fit to the low temperature data as described in the text.  The data 
for the parent compound is adapted from Ref.~\citen{Williams_16}.}
\end{center}
\end{figure*}

The inelastic triple-axis measurements at $\vec{Q}_{com}$~=~(1~0~0) are 
shown in Fig.~\ref{q_com}, at 30~K, above the transition (open circles), and 
at 4~K, below the transition (filled circles) for each of the measured 
dopings.  The data for the 1\% (panel (b)) and 5\% (panel (d)) samples were 
taken on the HB-1 spectrometer, which had a lower background than the same 
measurements on the HB-3 spectrometer for the other Fe-doped samples.  
However, all samples clearly show the opening of the gap in the excitation 
spectrum below the transition.  The same excitation in the parent compound is 
shown in Fig.~\ref{q_com}(a) for comparison (data adapted from 
Ref.~\citen{Williams_16}).  The solid line is a fit to the data, following the 
analysis of Ref.~\citen{Broholm_91} and Ref.~\citen{Williams_16}, given by:

\begin{eqnarray}
\tilde{I}({\bf Q},\omega)=
I & \left[ \frac{\hbar\gamma/\pi}{(\hbar\omega-\epsilon({\bf Q}))^2
	+(\hbar\gamma)^2} \right. \nonumber \\
& \left. -\frac{\hbar\gamma/\pi}{(\hbar\omega+\epsilon({\bf Q}))^2
	+(\hbar\gamma)^2} \right]
\label{eq1}
\end{eqnarray}

\noindent where $I$ is an overall scale factor for the intensity and 
$\hbar\gamma$ is the Half Width at Half Maximum (HWHM) for the Lorentzian 
functions.  With an energy gap $\Delta$, the dispersion relation reads:

\begin{equation}
\epsilon({\bf Q})=
\sqrt{\Delta^2+\hbar^2(\delta Q_{\perp}^2v_{\perp}^2+\delta 
	Q_{\parallel}^2v_{\parallel}^2)}
\label{eq2}
\end{equation}

\noindent where $\delta Q_{\perp,\parallel}=|({\bf Q}-{\bf 
Q}_0)_{\perp,\parallel}|$ is the projection of the difference of the wave 
vector transfer ${\bf Q}$ from the critical wave vector ${\bf Q}_0$ 
perpendicular and parallel, respectively, to the 
$\mathbf{\hat{c}}$-direction.  The velocities used were those of the parent 
compounds, where $v_H$~=~$v_K$~=~$v_{\perp}$~=~23.7(5)~meV$\cdot$\AA~and 
$v_L$~=~$v_{\parallel}$~=~32.5(7)~meV$\cdot$\AA~\cite{Williams_16}.  
Eq.~\ref{eq1} was multiplied by a Bose factor and convoluted with the 4D 
experimental resolution function using \textsc{Reslib}~\cite{Zheludev_07}.  
This under-estimates the elastic peak at (1~0~0) in Fig.~\ref{q_com} due to 
the elastic magnetic Bragg peak at this $\vec{Q}$, but more reliably 
reproduces the quasi-elastic signal at the incommensurate (1.4~0~0) in 
Fig.~\ref{q_inc}.  Since these measurements were most concerned with 
extracting the parameters of the inelastic excitation, no additional terms 
were included to model the elastic peak.  The values obtained from these fits 
are given in Table~\ref{fit_tbl}, below.

\begin{figure*}[tbh]
\begin{center}
\includegraphics[angle=0,width=\textwidth]{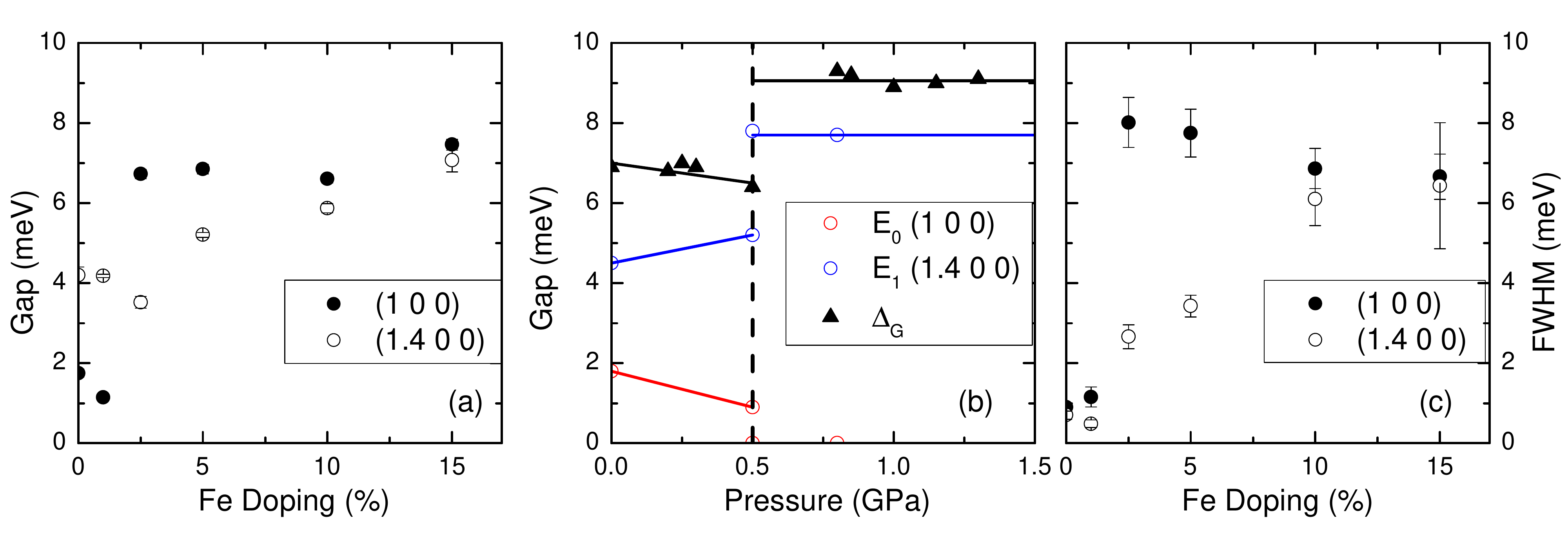}
\caption{\label{x_dep} (color online) (a) The gap at $\vec{Q}_{com}$ (filled 
circles) and $\vec{Q}_{inc}$ (open circles) as a function of Fe doping 
measured at T~=~4~K.  The values of the gap at 1\% doping are nearly unchanged 
from the parent compound.  Above 1\% doping, the gap at the commensurate 
wavevector increases dramatically, while the incommensurate gap increases 
continuously with Fe doping. (b) The value of the gap at $\vec{Q}_{com}$ (red 
circles), $\vec{Q}_{inc}$ (blue circles) and the gap measured by transport 
(black triangles) as a function pressure.  Figure reproduced with permission 
from Ref.~\citen{Bourdarot_10}, copyright American Physical Society. (c) The 
full width at half maximum (FWHM) of the excitations as a function of doping 
at T~=~4~K.  Similarly to the behavior of the gaps, the width of the 
excitations is nearly unchanged at 1\% doping.  Above 1\%, the width of the 
commensurate excitation is greatly increased, while the incommensurate 
excitation gradually broadens with increasing Fe doping.}
\end{center}
\end{figure*}

We see that in the 1\% doping, the commensurate excitation is nearly unchanged 
from the parent material; the gap and width are unchanged within error.  
However, we notice a dramatic change in the 2.5\% doped sample, where the 
excitation is substantially broadened in energy and is peaked at much higher 
energies.  The excitation is essentially unchanged with further increases in 
doping, with the gap energy and the width much larger than in the parent 
compound.  This trend is shown in Fig.~\ref{x_dep} where we can see the very 
abrupt changes in the gap (panel (a)) and the FWHM (panel (c)), which are 
relatively constant above 1\% doping.  It is also notable that the 
commensurate excitation is qualitatively unchanged across the phase 
transition, despite the emergence of the magnetic Bragg peaks at (1~0~0).  In 
agreement with the time of flight measurements, no other excitations are 
present in any of the samples.  

Fig.~\ref{q_inc} shows the excitation that is present below T$_0$ at 
$\vec{Q}_{inc}$~=~(1.4~0~0) as a function of doping.  This excitation was fit 
in the same manner as the commensurate excitation, shown by the solid lines in 
Fig.~\ref{q_inc}.  The values obtained from this fitting are given in 
Table~\ref{fit_tbl}.  As with the commensurate excitation, the incommensurate 
excitation shows very little change at 1\% doping relative to the parent 
compound.  However, above 1\% doping, rather than a discontinuous change, the 
incommensurate excitation exhibits a continuous broadening and upward shift in 
energy.  As with the doping dependence of the magnetic moment, the 
incommensurate excitation shows a discontinuous change from the hidden order 
to antiferromagnetic phases, as well as a continued evolution over the entire 
range of Fe doping.  This is apparent from looking at Fig.~\ref{x_dep}(a) and 
(c), where the gap and FWHM, respectively, show an increase over the full 
range of dopings measured.  The excitation appears to weaken continuously with 
increasing Fe doping, but is present in all dopings measured with no 
additional excitations present.

\begin{table}[htbp]
\resizebox{\columnwidth}{!}{
\begin{tabular}{|d|r|r|d|d|}
\hline
\multicolumn{1}{|c|}{Doping~} & \multicolumn{1}{c|}{~Wavevector~} & 
\multicolumn{1}{c|}{I (arb. units)} & \multicolumn{1}{c|}{$\Delta$ (meV)} & 
\multicolumn{1}{c|}{$\gamma$ (meV)}
\\
\hline
 & & & & \\
0.0\%~\cite{Williams_16} & (1~0~0)~~ & \multicolumn{1}{c|}{--} & 2.3(4) & 0.9(1) \\
1.0\%~ & (1~0~0)~~ & 1.55(3.77)~~ & 2.3(1) & 1.2(2) \\
2.5\%~ & (1~0~0)~~ & 6.99(2.13)~~ & 6.7(1) & 8.0(6) \\
5.0\%~ & (1~0~0)~~ & 10.28(3.08)~~ & 6.8(1) & 7.7(6) \\
10.0\%~ & (1~0~0)~~ & 7.01(1.95)~~ & 6.6(1) & 6.9(5) \\
15.0\%~ & (1~0~0)~~ & 6.04(1.34)~~ & 7.5(1) & 6.7(6) \\
 & & & & \\
0.0\%~\cite{Williams_16} & (1.4~0~0)~ & \multicolumn{1}{c|}{--} & 4.2(2) & 0.7(1) \\
1.0\%~ & (1.4~0~0)~ & 5.12(3.08)~~ & 4.18(4) & 0.48(9) \\
2.5\%~ & (1.4~0~0)~ & 5.26(2.26)~~ & 3.5(1) & 2.7(3) \\
5.0\%~ & (1.4~0~0)~ & 2.48(78)~~~ & 5.21(6) & 3.4(3) \\
10.0\%~ & (1.4~0~0)~ & 0.59(26)~~~ & 5.9(1) & 6.1(7) \\
15.0\%~ & (1.4~0~0)~ & 0.25(25)~~~ & 7.1(3) & 6.4(1.6) \\
\hline
\end{tabular}
}
\caption[]{Results of fitting the data in Fig.~\ref{q_com} and \ref{q_inc} to 
the Eq.~\ref{eq1}, as described in the text.  Data for the parent compound 
($x~=~0.0$) is taken from Ref.~\citen{Williams_16}.}
\label{fit_tbl}
\end{table}

Comparing these results to the gap measured by inelastic neutron scattering 
under pressure (shown in Fig.~\ref{x_dep}(b)), we see that there is a 
similarity when considering the incommensurate excitation (blue circles).  The 
application of pressure also increases the gap, though it is assumed that 
under pressure the gap jumps discontinuously at P$_0$~=~0.5~GPa and is 
constant above.  However, there may not be enough data points to be 
certain~\cite{Bourdarot_10,Williams_16,Villaume_08}.

\begin{figure}[tbh]
\begin{center}
\includegraphics[angle=0,width=\columnwidth]{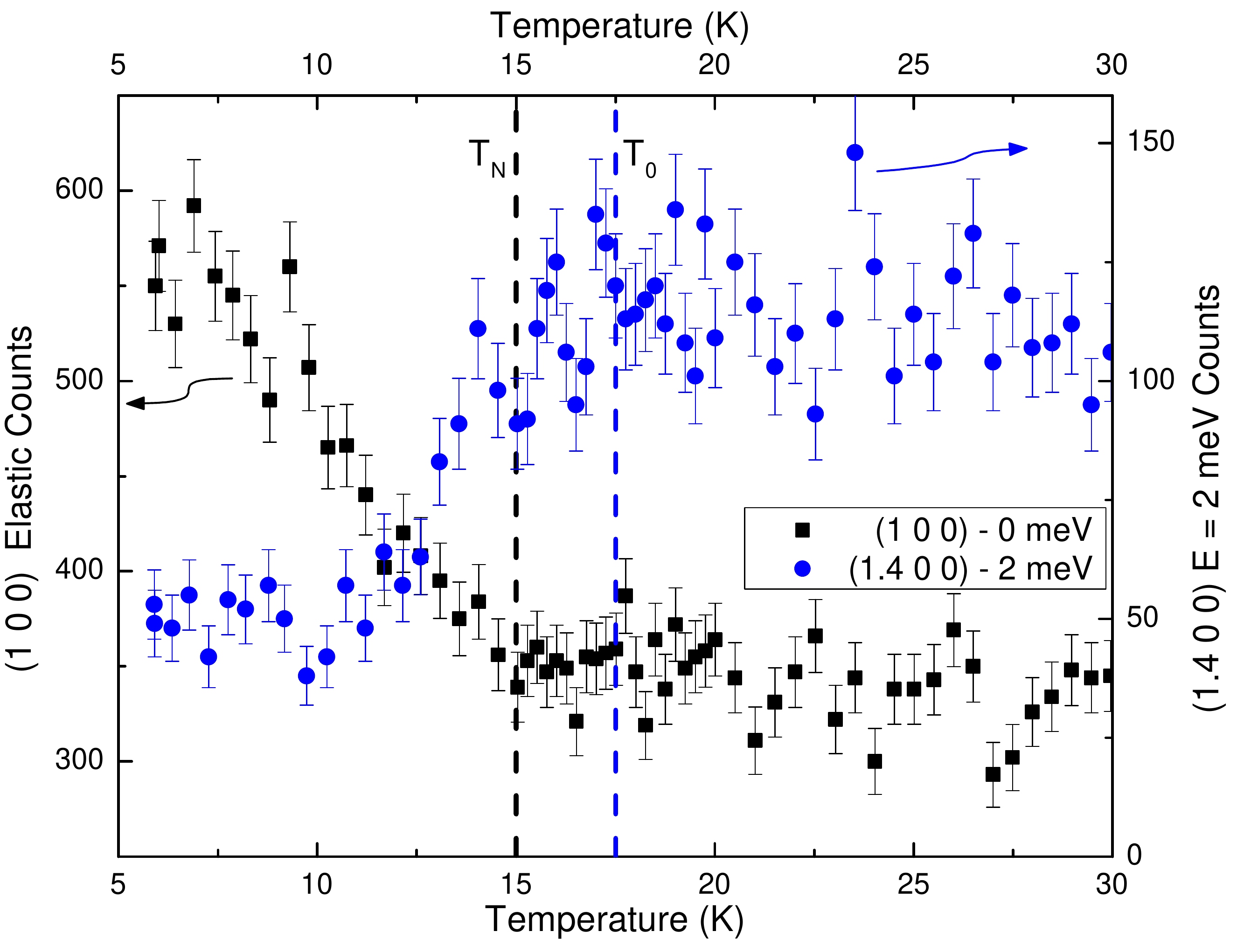}
\caption{\label{op} (color online) Plots of the order parameters for hidden 
order and antiferromagnetic phases for the 1\% Fe-doped sample.  The elastic 
magnetic Bragg peak (black squares) shows an onset around 15~K, coincident 
with the transition in the $\mu$SR measurements, while the opening of the gap 
at $\vec{Q}_{inc}$ (blue circles) onsets at 17.5~K, the same as for the parent 
compound and where the transition is seen by susceptibility~\cite{Wilson_16}.}
\end{center}
\end{figure}

\begin{figure}[tbh]
\begin{center}
\includegraphics[angle=0,width=\columnwidth]{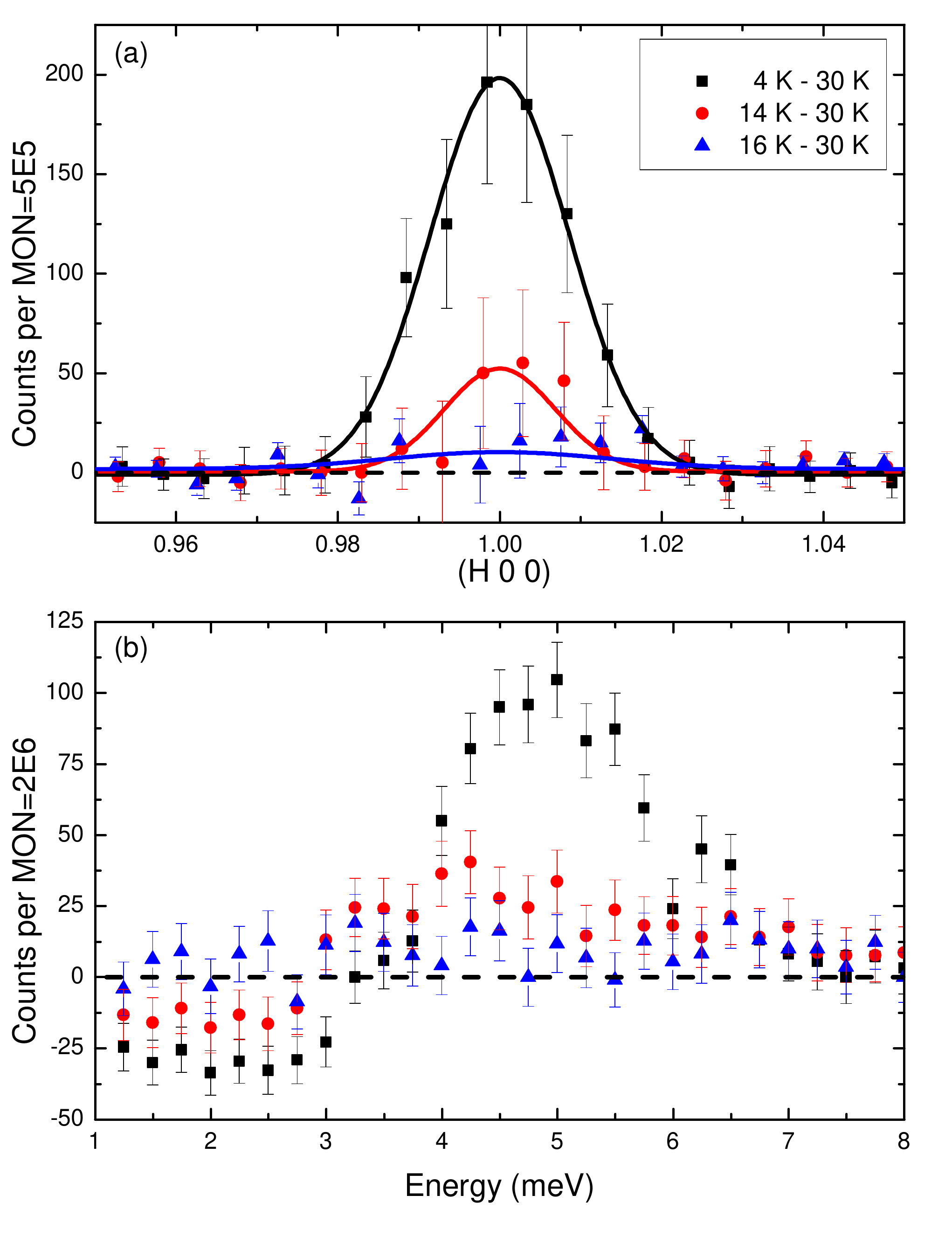}
\caption{\label{1pc} (color online) (a) The (1~0~0) magnetic Bragg peak 
shown at 4~K, 14~K and 16~K, subtracting the same data at 30~K.  Here we see 
the disappearance of the magnetic Bragg peak at a temperature below the hidden 
order transition at T$_0$~=~17.5~K.  (b) Energy scan at (1.4~0~0) at the same 
temperatures as in (a), showing the temperature evolution of the gap.  The gap 
is present at all temperatures, though the weak signal and small gap (within 
the experimental resolution) at 16~K make this less clear than the measurement 
shown in Fig.~\ref{op}.}
\end{center}
\end{figure}

Lastly, to more directly probe the relationship between the hidden order and 
the antiferromagnetic order, we measured the order parameters for both types 
of ordering simultaneously in the 1\% Fe doped sample, shown in 
Fig.~\ref{op}.  The black squares denote the peak intensity of the (1~0~0) 
elastic magnetic Bragg peak, while the blue circles are the scattering 
intensity at (1.4~0~0) and an energy transfer of 2~meV.  This shows the 
strength of the scattering at a point within the incommensurate gap, a 
measurement that was shown to determine the opening of the gap at T$_0$ in the 
parent compound~\cite{Wiebe_04}.  In agreement with the quantitative 
similarities of the excitations in the 1\% sample and the parent compound, as 
well as the bulk thermodynamic data~\cite{Kanchanavatee_11,Wilson_16}, we see 
the opening of the incommensurate gap at T$_0$~=~17.5~K.  However, in 
agreement with the $\mu$SR measurements~\cite{Wilson_16}, the onset of the 
antiferromagnetic order occurs at a slightly lower temperature, T$_N$~=~15~K.  
Despite the apparent variation in the transition temperatures, specific heat 
measurements see no entropy change between the hidden order and 
antiferromagnetic phases, emphasizing that the Fermi surface reconstruction 
happens at the upper transition~\cite{Kanchanavatee_11}. Recent magnetization 
and thermal expansion measurements also see evidence for the possibility of 
two transitions, though they suggest that this is also present at higher 
dopings ($\sim$5\%)~\cite{Ran_16}.  This may be due to variations in doping 
concentrations or a difference in sensitivity of the measurement techniques.

To verify the presence of two transitions, constant $\vec{Q}$ measurements 
were performed at 4~K, 14~K, 16~K, 18~K and 30~K to measure both the (1~0~0) 
magnetic Bragg peak and the opening of the gap at (1.4~0~0), shown in 
Fig.~\ref{1pc}.  It can be seen that the 14~K data shows a gap in the 
(1.4~0~0) excitation spectrum, and there is appreciable scattering at the 
(1~0~0) magnetic Bragg peak.  At 16~K, the magnetic Bragg peak is absent, 
within error, while the gap in the (1.4~0~0) constant-$\vec{Q}$ measurement 
had been reduced, it is still present.  Both measurements at 18~K are 
identical within error to the 30~K data.  This is consistent with the 
separation in temperature of the hidden order and magnetic transitions.

\section{\label{sec:level5}Discussion \& Conclusions}

We have presented a comprehensive set of elastic and inelastic neutron 
scattering measurements on a range of Fe-doped samples of 
U(Ru$_{1-x}$Fe$_x$)$_2$Si$_2$ with 0.01~$\le~x~\le$~0.15.  We have found that 
the onset of the antiferromagnetic phase occurs at very low doping, with the 
2.5\% doped sample showing an ordered moment of 0.51~$\mu_B$.  However, the 
1\% sample seems to show excitations that are nearly identical to the parent 
compound, but onsetting at a higher temperature than the antiferromagnetic 
moment. Combined with previous susceptibility and $\mu$SR 
measurements on these samples~\cite{Wilson_16}, there is strong evidence of 
different transition temperatures for the antiferromagnetic and hidden orders, 
in agreement with other techniques on different Fe-doped 
samples~\cite{Ran_16}.  Resistivity and specific heat measurements do not see 
any signatures of an abrupt phase transition between the hidden order and 
antiferromagnetic state~\cite{Kanchanavatee_11,Ran_16}.  This is consistent 
with no observed change in $\vec{Q}$ for the incommensurate excitation, which 
remains at the $\Sigma$-point of the hidden order phase, suggesting no change 
in the BZ between the antiferromagnetic and hidden order phases.  
Additionally, the $\mu$SR measurements see evidence for phase separation at 
low dopings, likely a result of the statistically-random distrubition of Fe 
dopants~\cite{Wilson_16}.  These dopings are also where the (1~0~0) magnetic 
Bragg peak does not show a rapid onset, seen in Fig.~\ref{hb1a}(b), which 
would be expected in samples with low doping concentrations.

All of the dopings that were measured show evidence for long-ranged magnetic 
order, with the moment size increasing as a function of doping.  This suggests 
that even far from the parent compound, there is still an evolution away from 
hidden order.  This increase in the magnetic moment is accompanied by a 
continuous increase in T$_N$, which peaks above the dopings studied at 
$\sim$40\% doping, before being suppressed to a paramagnetic state above 
$\sim$70\% doping.  Synthesis of large single crystals becomes difficult above 
15\% Fe doping~\cite{Kanchanavatee_11}, but $\mu$SR measurements up to 50\% Fe 
doping show that the magnetic moment decreases above 15\% Fe 
substitution~\cite{Wilson_16}.

The inelastic time-of-flight and triple-axis measurements show that both sets 
of excitations observed in the parent compound are present at all dopings 
measured.  However, while the excitations are qualitatively unchanged, there 
are dramatic changes in the quantitative properties above 1\% doping, most 
noticeably in the reduction of the intensity of the commensurate excitation.  
The increase in the gap and energy-broadening of the excitations at both the 
commensurate and incommensurate point occurs noticeably in the 2.5\% doped 
sample.  Both the magnitude of the gap ($\Delta$) and the width ($\gamma$) 
evolve continuously with doping, which is most apparent at the incommensurate 
point.  As observed with measurements of the parent compound under pressure, 
the increase in the gap at $\vec{Q}_{inc}$ coincides with an increase in 
T$_0$.  This also follows the monotonic increase in the magnetic moment with 
doping, suggesting that the critical doping is between 1\% and 2.5\%, but that 
the magnetic  moment and the excitations change continuously at higher dopings.

The pressure results have been somewhat unclear about the existence and 
properties of the commensurate excitation, with work performed at 0.62~GPa 
reporting its absence~\cite{Bourdarot_10,Villaume_08,Hassinger_10a}, while 
other work seeing a gap of $<$1~meV at 0.72~GPa~\cite{Aoki_09} and a gap of 
1.8~meV at 1.02~GPa~\cite{Williams_16}.  This has been interpreted as mode 
softening at the critical pressure, P$_C$~=~0.6~GPa, which may explain the 
changing value of the gaps as seen in the present case of Fe-doping.  However, 
the much larger gap and width in the Fe-doped samples clearly demonstrate that 
the behavior of the commensurate excitation under Fe doping is not the same as 
under applied pressure, which may suggest that the effect of Fe doping on the 
$Z$ point Fermi surface pocket is not strictly analogous to the changes that 
occur under hydrostatic pressure.  Furthermore, the change in the excitations 
point to evolutions in the Fermi surface with increasing Fe doping; this 
serves to increase the gap, suggesting that the Fermi surface pockets at the 
$\Sigma$, $Z$ and/or $\Gamma$ points distort slightly to change the optimal 
energy for the nesting.  This must occur without any Fermi surface 
reconstruction, as there is no entropy change across the HO-AF 
transition~\cite{Kanchanavatee_11}, nor do we see any change in the location 
of the incommensurate excitation ($\Sigma$), suggesting that the Fermi surface 
is not distorted in the antiferromagnetic state.  Drawing the analogy to the 
antiferromagnetic state induced by applied pressure, that transition similarly 
shows no Fermi surface reconstruction by quantum oscillation 
measurements~\cite{Hassinger_10b}.  We can make further comparison to the 
pressure-induced AF state by looking at the excitations seen by neutron 
scattering.  Under pressure, the gap at the incommensurate point similarly 
shows a slight increase, while the intensity of the excitations also 
increases~\cite{Williams_16}.  The intensity of the excitations does not 
increase with Fe doping, but this may be a result of impurities distorting the 
Fermi surface, serving to weaken the nesting that is undistorted in the case 
of applied pressure.  This can also be seen by comparing the width of the 
excitations, which are unchanged under pressure~\cite{Williams_16}, but 
dramatically broadened in the case of Fe doping.

This study serves to illustrate that URu$_2$Si$_2$ is ideally placed on the 
precipice of magnetic states: antiferromagnetism under pressure or Fe-doping, 
and even ferromagnetism under Re-doping~\cite{Butch_11}.  In all cases, we see 
that the excitation spectrum changes quantitatively, but not qualitatively, 
and is not destroyed by the emergence of the magnetically-ordered 
state~\cite{Williams_16,Williams_12}.  Thus this work demonstrates that in the 
Fe-doped compounds studied here, as with other perturbations, the hidden order 
state is not incompatible with magnetic order but rather that the electronic 
correlations are intimately related to magnetism.

\

\section{\label{sec:level6}Acknowledgments}

The authors would like to thank C.R.~Wiebe for helpful discussions as well as 
C.~Broholm for his input and collaboration on the parent 
compound~\cite{Williams_16}.  We also note that inelastic neutron scattering 
work on these compounds was submitted recently during the preparation of this 
manuscript~\cite{Butch_16} and we thank N.P.~Butch for sharing that work, 
whose results are consistent with the present study.

We acknowledge instrument support from S.~Chi, M.~Matsuda, L.M.~Debeer-Schmitt 
and D.~Pajerowski.  This research at ORNL's High Flux Isotope Reactor 
and Spallation Neutron Source was sponsored by the Scientific User Facilities 
Division, Office of Basic Energy Sciences, US Department of Energy.  Work at 
McMaster University was supported by the Natural Sciences and Engineering 
Research Council of Canada and the Canadian Foundation for Innovation.  T.J.W. 
acknowledges support from the Wigner Fellowship program at Oak Ridge National 
Laboratory.  M.N.W. acknowledges support from the Alexander Graham Bell Canada 
Graduate Scholarship program.

\end{document}